\documentclass[aps,showpacs,amsmath,amssymb,twocolumn,prx,longbibliography,superscriptaddress,notitlepage,floatfix,raggedbottom,10pt]{revtex4-2}
\usepackage[T1]{fontenc}
\usepackage[utf8]{inputenc}
\usepackage{color}
\usepackage{xcolor}
\usepackage{graphicx}
\usepackage{babel}
\usepackage{amsmath}
\usepackage{amsthm}
\usepackage{amsfonts}
\usepackage{mathrsfs}
\usepackage{dsfont}
\usepackage{bm}
\usepackage{appendix}
\usepackage{physics}
\usepackage{quantikz}
\usepackage[caption=false]{subfig}
\usepackage[
    unicode=true,
    pdfusetitle,
    bookmarks=true,
    bookmarksnumbered=false,
    bookmarksopen=false,
    breaklinks=true,
    pdfborder={0 0 0},
    pdfborderstyle={},
    backref=false,
    colorlinks=true]{hyperref}
\usepackage{placeins}
\usepackage{ascmac}

\begin{document}

\title{Error-mitigation aware benchmarking strategy for quantum optimization problems}

\author{Marine Demarty}
\affiliation{School of Informatics, University of Edinburgh, 10 Crichton Street, EH8 9AB Edinburgh, United Kingdom}

\author{Bo Yang}
\affiliation{LIP6, Sorbonne Universit\'e, CNRS, 4 place Jussieu, 75005 Paris, France}

\author{Kenza Hammam}
\affiliation{School of Informatics, University of Edinburgh, 10 Crichton Street, EH8 9AB Edinburgh, United Kingdom}

\author{Pauline Besserve}
\email{pauline.bess@proton.me}
\affiliation{School of Informatics, University of Edinburgh, 10 Crichton Street, EH8 9AB Edinburgh, United Kingdom}

\begin{abstract}
Assessing whether a noisy quantum device can potentially exhibit quantum advantage is essential for selecting practical quantum utility tasks that are not efficiently verifiable by classical means. For optimization, a prominent candidate for quantum advantage, entropy benchmarking provides insights based concomitantly on the specifics of the application and its implementation, as well as hardware noise.
However, such an approach still does not account for finite-shot effects or for quantum error mitigation (QEM), a key near-term error suppression strategy that reduces estimation bias at the cost of increased sampling overhead.
We address this limitation by developing a benchmarking framework that explicitly incorporates finite-shot statistics and the resource overhead induced by QEM.
Our framework quantifies quantum advantage through the confidence that an estimated energy lies within an interval defined by the best-known classical upper and lower bounds.
Using a proof-of-principle numerical study of the two-dimensional Fermi-Hubbard model at size $8\times8$, we demonstrate that the framework effectively identifies noise and shot-budget regimes in which the probabilistic error cancellation (PEC), a representative QEM method, is operationally advantageous, and potential quantum advantage is not hindered by finite-shot effects.
Overall, our approach equips end-users with a framework based on lightweight numerics for assessing potential practical quantum advantage in optimization on near-future quantum hardware, in light of the allocated shot budget.
\end{abstract}

\maketitle

\section{Introduction}

Recent years have witnessed remarkable progress in quantum hardware, together with advances in quantum error suppression techniques prior to full fault tolerance, which have led to a series of milestone experiments reporting quantum advantage on carefully designed benchmarking tasks~\cite{arute_quantum_2019, kim_evidence_2023, Abanin2025Observation}.
As quantum processors continue to scale and early fault-tolerant operations begin to be demonstrated~\cite{Hughes2025Trapped-ion, Ransford2025Helios, acharya_quantum_2025_compressed, Rosenfeld2025Magic}, there is growing anticipation that quantum advantage for real-world utility tasks becomes available in the near future, particularly in the quantum simulation of many-body systems beyond the practical capability of classical simulation~\cite{Lanes2025A, Huang2025The, Babbush2025The}.

The fact that those tasks would no longer be efficiently verifiable by classical means raises a central question: how can one assess whether a given quantum device is capable of providing quantum advantage for a target task in a principled and reliable manner?
Addressing this question calls for benchmarking frameworks that go beyond hardware-level performance metrics and instead assess quantum advantage at the level of the application itself.
Namely, such benchmarks should remain agnostic to specific hardware implementations and algorithmic details, while capturing the fundamental limitations imposed by noise on the achievable quality of outputs.

\begin{figure}[htbp]
    \centering
    \includegraphics[width=0.85\linewidth]{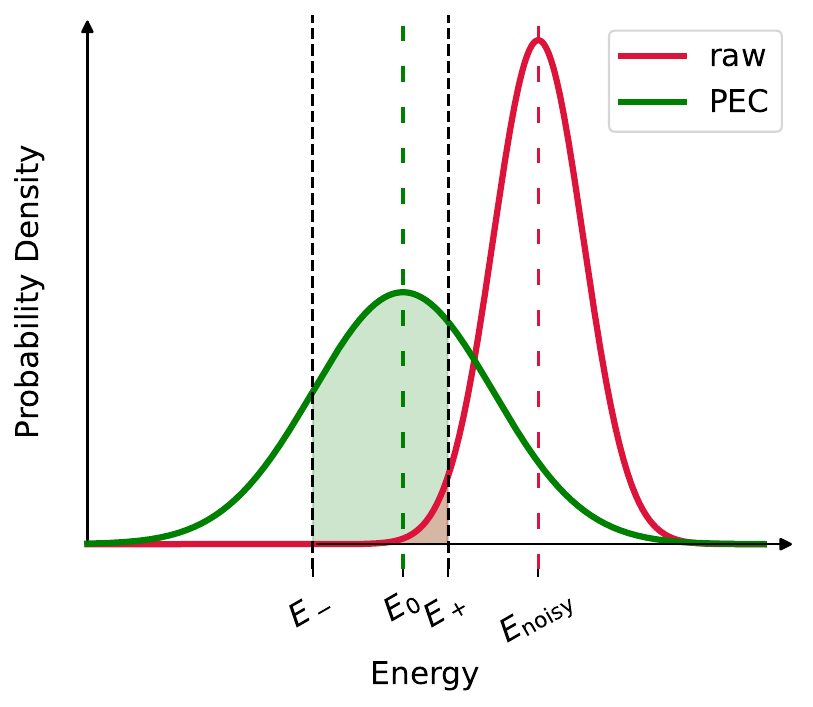}
    \caption{
        \textbf{Illustration of the method.} 
        For a given number of shots, compared with the raw results in red, probabilistic error cancellation (PEC) yields an unbiased energy distribution with larger variance.
        We consider the error-mitigated experiment to have succeeded if we obtain an estimated energy within the interval $[E_{-}, E_{+}]$, where $ E _ {-} $ and $ E _ {+} $ are the tightest lower and upper bounds to the target energy achieved by classical algorithms.
        The success probability $\mathbb{P}_\text{success}$ is then depicted as the area of the green-shaded region between these bounds.
        The success probability for unmitigated distribution (red-shaded area) can be used to assess whether applying QEM is advantageous.
    }
    \label{fig:method}
\end{figure}

In this direction, the entropy benchmarking framework~\cite{stilck_franca_limitations_2021_corrected, demarty_entropy_2024, Besserve2025Probing} has emerged as a powerful approach, providing a rigorous criterion that links entropy accumulation on a quantum processor to performance degradation in optimization problems such as ground-state energy estimation in physical and chemical models.
Specifically, entropy benchmarking relies on efficiently computable lower bounds to the Gibbs-state energy-entropy boundary to determine, from the physical entropy of a prepared state, whether potential quantum advantage for a given task remains achievable.
This framework was recently rendered practical and applied to the Fermi-Hubbard ground-state problem that covers a vast class of physical models of interest, yielding conservative no-go results for near-term quantum devices~\cite{Besserve2025Probing}.

However, existing benchmarking methods, including entropy benchmarking, are not yet fully applicable to realistic quantum experiments operating under finite-shot constraints.
In particular, they do not account for quantum error mitigation (QEM)~\cite{li_efficient_2017, temme_error_2017, endo_practical_2018, yang2022efficient, cai_practical_2023}, a central near-term error-suppression technique in which estimation bias is traded for increased sampling overhead. 
Because QEM reconstructs clean expectation values corresponding to an effectively noiseless ansatz via classical post-processing, the physical state entropy no longer reflects the quality of the error-mitigated estimate.
As a result, existing entropy benchmarking with deterministic entropy-energy curves falls short of characterizing the end-to-end performance of error-mitigated quantum computations with a statistical nature induced by finite shot counts.

Moreover, a complementary challenge in benchmarking also arises on the QEM side: there is no quantitative, task-level criterion to determine whether QEM ultimately enhances or undermines quantum advantage under a fixed shot budget.
QEM methods are known to incur exponential sampling overhead~\cite{Takagi2022Fundamental, takagi_universal_2023, tsubouchi_universal_2023, quek_exponentially_2024}, whereas it is widely believed that there is a finite regime of device scale and noise strength in which QEM remains practically feasible~\cite{Zimboras2025Myths}.
This underscores the strong demand for a principled benchmarking framework that can operationally and quantitatively determine whether a given quantum resource is in a regime in which applying QEM is advantageous for deploying the target quantum utility task.

To address these demands, we develop a quantum-advantage benchmarking framework that explicitly addresses finite-shot statistics and the sampling overhead induced by QEM, while retaining the algorithm- and hardware-agnostic spirit of entropy-based approaches.
Instead of deterministically mapping state entropy to the lower bound of its corresponding Gibbs-state energy, our framework evaluates the confidence with which the estimated energy lies within a classically certified interval that contains the true ground-state energy (see Fig.~\ref{fig:method}), a scenario we define as quantum advantage.
This enables a quantitative, shot-noise-aware assessment of potential quantum advantage under realistic execution constraints given a fixed shot budget and the device's underlying physical noise level. 
From the QEM perspective, our framework also provides a concrete, task-level criterion to identify the regimes in which QEM can be operationally beneficial.

To illustrate the framework, we consider ground-state energy estimation for the two-dimensional Fermi-Hubbard model and adopt probabilistic error cancellation (PEC)~\cite{temme_error_2017, endo_practical_2018} as a representative QEM strategy.
Through a proof-of-principle numerical study, we compare the success probabilities with and without error mitigation and identify regimes in which applying PEC becomes advantageous under realistic noise levels and finite shot budgets. 
These results demonstrate that the proposed framework can delineate when QEM meaningfully enhances the prospects of achieving quantum advantage, thereby providing a practical decision criterion for deploying QEM in near-term quantum utility tasks.

\FloatBarrier

\section{Methodology}

In this section, we focus on a benchmarking strategy for ground state energy estimation of lattice Hamiltonians based on PEC, an ubiquitous error mitigation method presented in Appendix \ref{app:pec}. The goal is to come up with a simple strategy to take QEM into account when assessing the quality of the result stemming from a quantum computing approach. We consider the problem of estimating the ground state energy $E_0$ of a Hamiltonian $\hat{H}$: $E_0 = \min_{\rho} \mathrm{Tr}(\rho \hat{H}).$

\paragraph{\textbf{Connection between PEC distribution variance and sampling overhead}}
The starting point of the analysis is the connection between the negativity, usually referred to as the sampling overhead, and the precision-shots trade-off at work within a chosen PEC strategy. Let us consider the Pauli decomposition of the Hamiltonian observable and denote by $\lVert H\rVert_2^2$ the squared norm-2 of the Pauli weights, which does not depend on the fermion-to-qubit encoding. Along with characteristics of the noise to be counteracted with PEC, it allows to define a minimum number of shots for the PEC procedure to reach a certain precision. Let indeed $\gamma$ be the negativity associated with the quasi-probability distribution chosen to combat the noise in one layer of a $D$-layer ansatz. Then, to get a $\sigma^2$-variance estimate of $H$ over a circuit instance using Pauli averaging, one needs to allocate a number of shots~\cite{tsubouchi_universal_2023} 
\begin{equation}
    N_{\mathrm{shots}} \geq \frac{\lVert H \rVert_2^2}{\sigma^2} \beta \gamma^{2D}
    \label{eq:tsubouchi}
\end{equation}
where $\beta$ depends on the (layerwise) noise channel. Note that for depolarizing noise, $\beta=1$.

\paragraph{\textbf{Definition of the `success' domain of PEC}}
Second, we formalize what a successful ground state energy estimation aided by PEC is. It has been shown that, assuming noise was characterized and inverted perfectly, PEC yielded a bias-free estimate of the observable at the expense of an increased variance compared with raw estimation. As a consequence, we can make use of Relation \ref{eq:tsubouchi} to assess whether or not within a given shot budget and a given circuit the PEC strategy is likely to output a result that is better than the state-of-the-art classical result.

We assume we have a circuit that, in the absence of noise, would prepare the target ground state.
Such an assumption enables us to consider the effect of quantum and shot noise separately from the problem of expressivity and/or circuit optimization. 
We consider noisy executions of this circuit estimating $E_0$, along with noisy executions of the modified circuits called for by the implemented PEC procedure. 
Let us assume as well that we have, from classical methods, both an upper and a lower bounds to the true ground state energy, denoted respectively as $E_+$ and $E_-$. Upper bounds are ubiquitous: for instance, they are given by variational optimization within a subspace of physical states. For lattice Hamiltonians, lower bounds may be obtained from geometrical arguments as observed by Anderson ~\cite{anderson_limits_1951} and further expanded in~\cite{valenti_rigorous_1991} and~\cite{eisert_note_2023}. 

We further make the assumption that the distribution of the energy observable within PEC is a normal distribution centered on $E_0$. This should be a reasonable assumption provided we consider a number of samples which is high enough, and assuming a very fine characterization of the noise. 

To take into account the statistical aspect of energy estimation, we define the ground state estimation task to have been successful on the quantum computer with PEC over classical methods if we get an estimation which falls between the best classical bounds $E_-$ and $E_+$ with sufficiently high probability (confidence) $\mathbb{P}_{\mathrm{success}} = 1-\delta$. In other words, we define the success probability from the probability measure of the interval $(E_-, E_+)$ defined from known bounds to the ground state energy, as represented in green on Fig.~\ref{fig:method}, and we define quantum advantage as passing a pre-defined threshold in terms of this success probability. Properties of normal distributions, recalled in Appendix \ref{app:gaussian_distribs}, allow for such a success probability to be linked to defining parameters of the distribution.

\paragraph{\textbf{Expression for the confidence in quantum advantage in the presence of PEC}}
Since we assumed the energy distribution yielded by PEC to be the normal distribution $\mathcal{N}(E_0, \sigma^2)$, this fraction of the probability weight of density denoted by $p(x)$ can be measured - albeit not analytically- by means of the error function. Namely, we can show that:

\begin{align}
    \mathbb{P}_{\mathrm{success}} &= \int \limits_{E_-}^{E_+} p(x)dx \nonumber\\
    &= 1-\mathbb{P}(X \geq E_+) - \mathbb{P}(X \leq E_-) \nonumber\\
    &= \frac{1}{2} \left[ \mathrm{erf}\left(\frac{E_+-E_0}{\sigma \sqrt{2}}\right) - \mathrm{erf}\left(\frac{E_--E_0}{\sigma \sqrt{2}}\right) \right]. \label{eq:P_success_general}
\end{align}

Replacing $\sigma$ with its best-case value given by \ref{eq:tsubouchi} for global depolarizing noise with total sampling overhead $\gamma_{\mathrm{tot}}$, we obtain

\begin{align}
    \mathbb{P}_{\mathrm{success}} =  &\frac{1}{2} \left[ 
        \mathrm{erf}\!\left(\left(E_+-E_0\right)\frac{\sqrt{N_{\mathrm{shots}}/2}}{\gamma_{\mathrm{tot}}\lVert H\rVert_2}\right) \right. \nonumber \\
    &\left. - \mathrm{erf}\!\left( \left(E_--E_0\right)\frac{\sqrt{N_{\mathrm{shots}}/2}}{\gamma_{\mathrm{tot}}\lVert H\rVert_2}\right) \right].
    \label{eq:P_success}
\end{align}

\paragraph{\textbf{An approximation to the confidence}}
Since we do not know $E_0$, Equation \ref{eq:P_success} does not provide a directly actionable formula for the success probability. We can however consider the error-mitigated distribution whose mean is $(E_-+E_+)/2$, the center of the interval defined by known classical bounds to $E_0$. Since the error function is an odd function, we get

\begin{equation}
    \widetilde{\mathbb{P}}_{\mathrm{success}} = \mathrm{erf}\left(\left(\frac{E_+-E_-}{2}\right)\frac{\sqrt{N_{\mathrm{shots}}/2}}{\gamma_{\mathrm{tot}}\lVert H \rVert_2}\right)
    \label{eq:P_tilde}.
\end{equation}

We provide numerical evidence that this is a reasonable assumption in Appendix \ref{app:effect_error_centering}, in that the proxy value $\widetilde{\mathbb{P}}_{\mathrm{success}}$ is likely to be close to $\mathbb{P}_{\mathrm{success}}$.

\paragraph{\textbf{Confidence in the success of the non-error-mitigated strategy}}
In the case PEC or any other QEM method is \emph{not} resorted to, we get upon measuring the Hamiltonian's expectation value an energy distribution that is biased. However, given a shot budget, this distribution exhibits a smaller variance than the PEC distribution. As a consequence, even before quantum advantage can be envisioned, there is a notion of break-even for PEC, just as is the case in error-corrected schemes. This motivates the comparison between the PEC strategy and the strategy based on raw results, where we can readily define a similar success probability. Indeed, in the presence of depolarizing noise, the bias $E_\text{noisy} - E_0$ can be readily computed from the depolarization probability (see Appendix \ref{app:bias_depol}) and Equation \ref{eq:P_success} can be adapted into

\begin{align}
    \mathbb{P}_{\substack{\text{success}, \\ \,\text{raw}}} &= \frac{1}{2} \left[ \mathrm{erf}\left(\frac{E_+-E_\text{noisy}}{\sigma_\text{noisy} \sqrt{2}}\right) \right. \nonumber \\ 
    & \left. - \mathrm{erf}\left(\frac{E_--E_\text{noisy}}{\sigma_\text{noisy} \sqrt{2}}\right) \right]\,, \label{eq:P_success_raw}
\end{align}

where
\begin{equation}
    \sigma_\text{noisy} = \frac{||H||_2}{\sqrt{N_\text{shots}}}
\end{equation}
is the best-case standard deviation of the energy distribution without error mitigation (from Eq.~\ref{eq:tsubouchi}) and $E_\text{noisy}$ is its mean, which is given by Eq.~\eqref{eq:E_noisy}. We use the same approximation as in Eq.~\ref{eq:P_tilde}, namely $\displaystyle E_0 \leftarrow \frac{E_- + E_+}{2}$.
This probability is represented with a red shade on Fig.~\ref{fig:method}. 

We underline however that, due to the bias, the definition of quantum advantage that we use in this case is more debatable from an information-theoretic perspective as in the presence of a strong bias, we will deem the success probability to be higher for a more spread distribution than a peaked one.

\section{Results}
\begin{figure}
     \centering
     \subfloat[Error-mitigated (PEC)\label{subfig:success_prob_PEC_globalDP}]{
        \includegraphics[width=\linewidth]{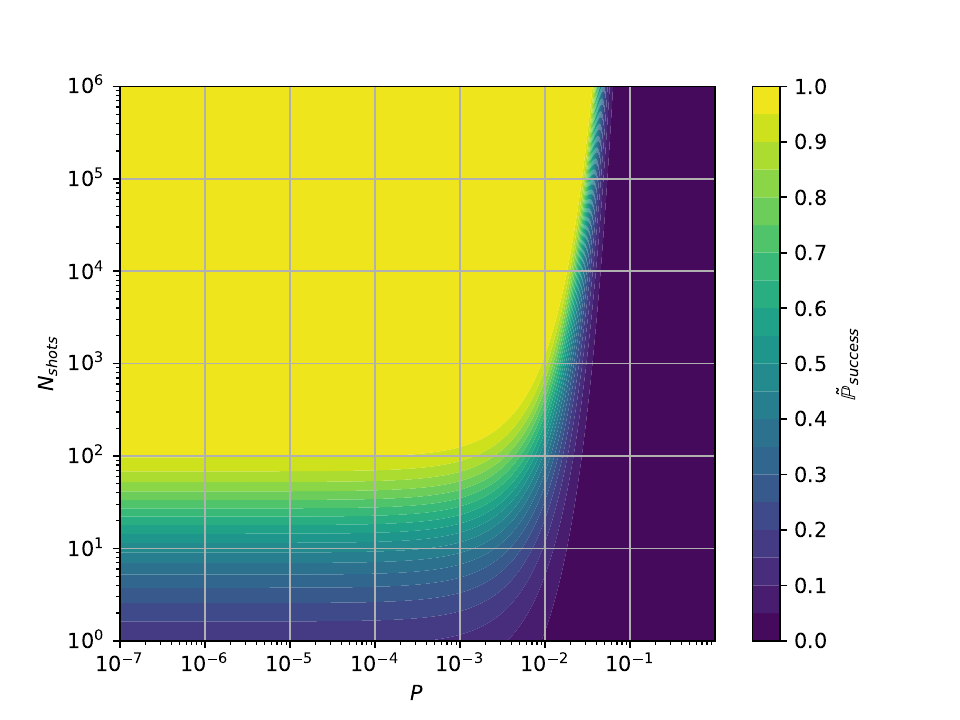}
     }
     
     \hfill
     \subfloat[Raw\label{subfig:success_prob_noPEC_globalDP}]{
        \includegraphics[width=\linewidth]{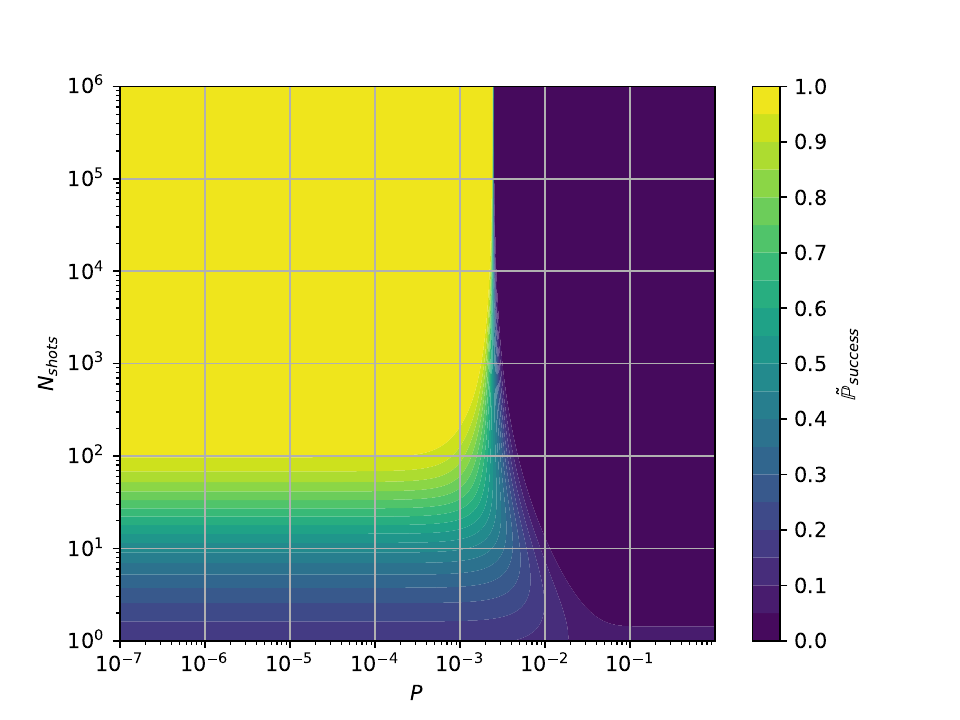}
     }
    \caption{
        \textbf{Success probability of ground state energy estimation under global depolarizing noise.} We consider a 2D Fermi-Hubbard Hamiltonian with parameters $(U, t, \mu) = (8, 1, 3.75)$ with $L=64$ sites on a square lattice. We assume the candidate ground state was obtained with a layered circuit with width $n=2L=128$ and depth $D=L=64$ under global depolarizing noise with layerwise global depolarizing probability $P$. We take values of the lower and upper bounds to the ground state energy for this specific energy minimization problem from~\cite{valenti_rigorous_1991}. Figures \ref{fig:success_prob_globalDP}(a) and \ref{fig:success_prob_globalDP}(b) show the success probability in the presence \eqref{eq:P_tilde} and absence \eqref{eq:P_success_raw} of PEC respectively, as a function of the noise level $P$ and the available number of shots $N_\text{shots}$.
    }
    \label{fig:success_prob_globalDP}
\end{figure}

To illustrate our method, we consider the 2D Fermi-Hubbard Hamiltonian~\cite{hubbard_electron_1997}, a model of fermions on a lattice which reads 
\begin{equation}
    \hat{H} = -t \sum_{\langle i,j \rangle, \sigma} \hat{c}_{i\sigma}^\dagger \hat{c}_{j\sigma} + U \sum_i \hat{n}_{i\uparrow} \hat{n}_{i\downarrow} - \mu \hat{N}\,,
\end{equation}
where $\hat{c}^\dagger_{i\sigma}, \hat{c}_{i\sigma}$ are the fermionic creation/annihilation operators, $\langle i,j \rangle$ refers to sites $i$ and $j$ which are connected in the lattice, $\hat{n}_{i\sigma} = \hat{c}_{i\sigma}^\dagger \hat{c}_{i\sigma}$ and $\hat{N} = \sum_{i, \sigma} \hat{n}_{i\sigma}$ is the number operator. This model is parametrized by the hopping parameter $t$, the onsite Coulomb interaction $U$ and the chemical potential $\mu$.

As a foundational model for strongly correlated systems, the Fermi-Hubbard model enables the exploration of rich phase diagrams that inform the design of new materials across diverse applications \cite{zhong_electronics_2015, haddad_review_2022}. Since many materials exhibit a layered structure that is amenable to a two-dimensional description, the two-dimensional Fermi-Hubbard model has been thoroughly studied ~\cite{leblanc_solutions_2015}, providing a range of classical results against which to benchmark. Yet, it becomes intractable for large system sizes, making it a prominent candidate for quantum advantage~\cite{fauseweh_quantum_2024}.

As an example, we consider a square $8 \times 8$ lattice ($L=64$ sites) with periodic boundary conditions, a size similar to what has been studied in Ref.~\cite{Besserve2025Probing}. We choose parameters $t=1$, $U=8$ and $\mu=3.75$, for which classical results are readily available for the benchmark. In particular, Ref.~\cite{valenti_rigorous_1991} provides, in the thermodynamic limit, a lower bound to the ground state energy density per site $E_{-}/(Lt) = -4.544$ by means of cluster decomposition. It additionally references the energy density per site $E_{+}/(Lt) = -3.8365$ stemming from the so-called Gutzwiller approximation and usually considered to be an upper bound to the ground state energy.

The state of this system may be represented for instance with a quantum register comprising $n=2L=128$ qubits using the Jordan-Wigner transformation~\cite{jordan1928paulische, somma2002simulating}.
To build the ground state of such a Hamiltonian with a quantum circuit, we assume we use the Hamiltonian Variational Ansatz circuit (HVA \cite{wecker_towards_2015}). For such an ansatz circuit, it is expected from evidence at small scale that the number of circuit layers must scale linearly with the number of sites $L$ for the circuit to be expressive enough and hope to build the ground state~\cite{cade_strategies_2019}. Therefore, we take the depth of the quantum circuit to be $D=L=64$.
We assume that each layer of the quantum circuit is followed by a layer of global depolarizing noise with depolarizing probability $P$. For such a circuit, the optimal overhead factor in the sampling cost of PEC or total negativity can be shown to be~\cite{takagi_optimal_2021}
\begin{equation}
    \gamma_\text{tot} = \left(\frac{1+(1-\frac{2}{d^2})P}{1-P}\right)^D\,,\label{eq:gamma_tot_globalDP}
\end{equation}
with $d=2^n$ the dimension. Substituting the total negativity in Eq.~\eqref{eq:P_tilde} with this expression, we can plot the success probability as a function of the noise level $P$ and the available number of shots $N_\text{shots}$, as represented in Fig.~\ref{fig:success_prob_globalDP}(a).

We observe that, to achieve a given success probability (with equal probability being represented by the grey curves delineating different hues in Fig.~\ref{fig:success_prob_globalDP}(a)), the number of shots $N_\text{shots}$ scales exponentially with the noise level $P$. This was expected as fixing the success probability means fixing the ratio $\sqrt{N_\text{shots}}/\gamma_\text{tot}$, which in turns means that the number of shots must satisfy
\begin{align}
    N_\text{shots} &\propto \gamma_\text{tot}^2 = \left( \frac{1+(1-\frac{2}{d^2})P}{1-P}\right)^{2D}\,,
\end{align}
which in the small noise limit ($P$ sufficiently small) reads
\begin{align}
    N_\text{shots} &\propto e^{4(1-\frac{1}{d^2}) D P}\,.
\end{align}

For comparison with the error-mitigated success probability, we also compute the success probability when PEC is not used, meaning raw energy estimation is performed on the noisy circuit.
We plot the resulting probability of quantum advantage as a function of the available number of shots $N_\text{shots}$ and the noise level $P$ in Fig.~\ref{fig:success_prob_globalDP}(b). The evolution of the success probability in the absence of QEM is more complex than in the presence of PEC: this is because, unlike in the PEC case, the distribution of the non-error-mitigated energy estimate is centered on a deviated energy value (mean) depending on the noise level $P$. This means that the noise level now influences the distribution and, therefore, the success probability in two ways: via the variance which depends on the negativity and thus on $P$, but also via the mean.

In the low noise level regime (small $P$; left part of the figure), both the PEC and the raw energy distributions should be similar since the sampling overhead $\gamma_\text{tot}$ of PEC is negligible ($\gamma_\text{tot} \sim 1$) and so is the shift of the mean in the raw case ($E_\text{noisy} \sim E_0$). This can be seen by the left part of Figs.~\ref{fig:success_prob_globalDP}(a) and \ref{fig:success_prob_globalDP}(b) being identical.

For larger noise levels (larger $P$), and for $N_\text{shots}$ sufficiently large, we notice in the raw energy estimation scenario (Fig.~\ref{fig:success_prob_globalDP}(b)) the presence of a vertical boundary where the success probability suddenly goes from $1$ to $0$ for some depolarizing probability $P$. Intuitively, for a large number of shots, the energy distribution is peaked and this sudden drop in success probability arises as soon as the mean value is equal to the upper bound to the ground state energy, $E_\text{noisy} = E_{+}$. Rewriting this condition as a condition on the depolarizing probability, we obtaim
\begin{equation}
    P = 1 - \left( \frac{E_{+} - \frac{\operatorname{Tr}\left[\hat{H}\right]}{d}}{E_{0} - \frac{\operatorname{Tr}\left[\hat{H}\right]}{d}}\right)^{\frac{1}{D}}\,,
\end{equation}
which for the chosen example corresponds to a threshold value $P=2.4\times 10^{-3}$, as observed.

On the lower right portion of Fig.~\ref{fig:success_prob_globalDP}(b), for a fixed noise level $P$ larger than the above threshold, we see that increasing the number of shots decreases the success probability instead of increasing it unlike in Fig.~\ref{fig:success_prob_globalDP}(a). This is because beyond this probability $P$, the center of the energy distribution is outside the success area defined by the known bounds $E_{-}$ and $E_{+}$. Since increasing the number of shots makes the distribution more peaked, more probability weight lies outside the success area: the success probability decreases. This should be understood as an artefact of the chosen metric, rather than a proper result.

\begin{figure}
    \centering
    \includegraphics[width=0.9\linewidth]{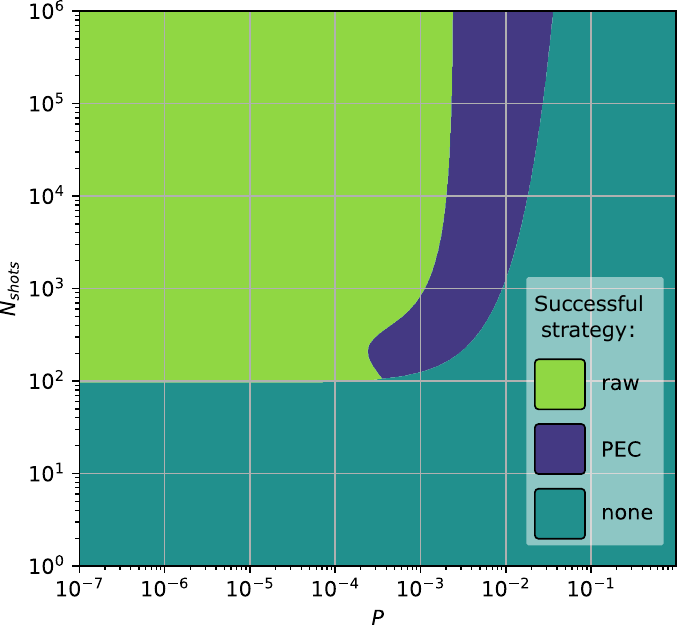}
    \caption{\textbf{Winning strategy as a function of the layerwise global depolarizing probability and the shot count.} We consider a 2D Fermi-Hubbard Hamiltonian with parameters $(U, t, \mu) = (8, 1, 3.75)$ with $L=64$ sites. We assume the quantum solution for the ground state energy was obtained with a layered circuit with width $n=2L=128$ and depth $D=L=64$ under global depolarizing noise with layerwise global depolarizing probability $P$. We take lower bounds and upper bounds to the ground state energy for this specific energy minimization problem from~\cite{valenti_rigorous_1991}. We show three regimes as a function of the noise level $P$ and the available number of shots $N_\text{shots}$ and for a fixed threshold value of $0.95$: `PEC' means PEC should be preferred over the raw distribution and ensures that the success probability is above the threshold, `raw' means the raw distribution should be preferred over PEC and gives a success probability above the threshold, and `none' corresponds to the regime where the success probability of both strategies is below the threshold.}
    \label{fig:three_regimes_winning_strategy_globalDP}
\end{figure}

Having access to the success probabilities in the presence and absence of PEC enables us to readily plot some type of phase diagram with three regions based on the noise level $P$ and the allocated number of shots $N_\text{shots}$ in Fig.~\ref{fig:three_regimes_winning_strategy_globalDP}. To this end, we define quantum advantage as the scenario in which $\widetilde{\mathbb{P}}_\text{success}$ is above a (somewhat arbitrarily) fixed threshold, which is set here to $\widetilde{\mathbb{P}}^\text{(QA)}_\text{success}=0.95$, and monitor for which regime this threshold is attained.

The first regime called `raw' is the case where the success probability is higher without PEC than with PEC and above the threshold; this is the regime where PEC is not required and quantum advantage is achieved with high probability. The second regime called `PEC' is the case where PEC performs better than in the absence of PEC, and its success probability is above the threshold; this is the regime where PEC should be used and quantum advantage is achieved with high probability. As a convention, for equal success probabilities (three matching decimals), we consider the raw strategy to be winning over PEC as it is less costly to implement than PEC. Lastly, the third `none' region corresponds to the case where both in the absence and in the presence of PEC, the success probability is below the threshold of $0.95$; we deem that no strategy is considered successful as none guarantees quantum advantage with sufficiently high probability.

We observe that the $(P, N_\text{shots})$ space is divided into three main regions. For low number of shots $N_\text{shots}$ (below around $10^{2}$ here), quantum advantage cannot be guaranteed with high probability ($>95\%$) regardless of whether PEC is used. This is because the probability distribution in both cases is too spread out and a non negligible portion of it thus lies outside the success area defined by the known energy bounds. 

In the low noise regime (typically $P<10^{-4}$ here), both the `PEC' and `raw' distributions have most of their probability weight in between the classical bounds, and so both success probabilities are close to 1. Based on the chosen convention, this explains why for $N_\text{shots}$ sufficiently high and in this low noise regime, using raw results is the winning strategy.

As the noise increases and for a fixed shot budget $N_\text{shots}>10^2$, the raw distribution shifts slowly outside of the success area due to an increasing bias, but it keeps a constant shape (variance), meaning the success probability keeps decreasing until it is below the $0.95$ limit. On the other hand, the PEC distribution stays centered in the middle of the success area but becomes more spread out as $P$ increases, meaning the success probability decreases as well although at a different speed than the raw distribution. As a consequence, there exists a region in terms of $P$ for which PEC is advantageous. Then, for larger $P$, neither the raw distribution nor the error-mitigated one guarantees quantum advantage with the required probability. 

It is easy to see that, given access to the calibration data of a quantum computer 
such as typical gate fidelities, and provided the circuit has the required structure to enable the preparation of the target state, one can easily predict from such a diagram (i) whether quantum advantage is achievable with high probability and (ii) if so, whether or not to deploy PEC.

For instance, for a state-of-the-art quantum computer with two-qubit gate error around $p=3.10^{-4}$~\cite{loschnauer_scalable_2024} and assuming all-to-all connectivity, the $N_\text{2Q/layer}\sim 2L = 128 $  two-qubit gates per HVA layer translate into a layerwise global depolarizing probability $P=1-(1-p)^{N_\text{2Q/layer}} \sim 4.10^{-2}$, which is the boundary of the PEC region for the maximum shot count investigated, $N_\text{shots} = 10^6$. On the other hand, assuming an improvement of one order of magnitude, we get $P \sim 4.10^{-3}$, a value for which PEC is successful for a modest number of shots around $10^3$. Although our analysis is based on strong assumptions and does not take into account any residual bias in PEC, it tends to demonstrate that hardware improvements could lead to a 'Goldilocks zone', where a PEC-enabled useful quantum advantage could be found.
This provides a novel insight, contrasting with negative results regarding the scaling of PEC~\cite{quek_exponentially_2024}.

\section{Conclusion and future directions}

In this work, we introduced a QEM-aware benchmarking framework for quantum optimization that is explicitly designed for realistic finite-shot experiments, where QEM plays a central role by trading estimation bias for increased sampling overhead.
The key implication is that once error mitigation is applied, end-to-end performance becomes intrinsically statistical.

Accordingly, our framework consists in evaluating the likeliness of quantum advantage as the confidence (success probability) that the estimated energy lies within known classical upper and lower energy bounds containing the true ground state energy. 
Using theoretical results for the widely studied probabilistic error cancellation (PEC) technique and assuming a bias-free normal distribution for the resulting energy estimator, we obtain a numerically assessable expression for the success probability, explicitly depending on the noise level $P$, the circuit depth $D$ and the shot budget $N_\text{shots}$.
Illustrated on a $8 \times 8$ square lattice Fermi-Hubbard model instance under global depolarizing noise, the proposed benchmarking method enables to partition the $(P, N_\text{shots})$ space into regimes where PEC is preferred and improves the prospects for quantum advantage over raw sampling, regimes where PEC is not required, and regimes where neither strategy guarantees quantum advantage with a sufficiently high probability defined by a set threshold.
The main takeaway of our work is thus a task-level criterion that captures the interplay between shot count and variance via the negativity dependence of PEC, rather than relying on a deterministic energy-entropy picture. 

Although we analyze a particular Hamiltonian model instance, noise model and ansatz circuit, our framework can be applied to other lattice Hamiltonians for which such classical lower and upper bounds are available, as well as to other noise models and to other ansatz circuits.

Moreover, a natural extension of this work would be to formulate straightforwardly analogous success probability benchmarks for other prominent QEM protocols reported in the literature~\cite{Li2017Efficient, koczor_exponential_2020, mcclean2021, Yoshioka2022Generalized, Yang2025Resource-efficient}, and to assess how the resulting regimes change under more realistic noise and error mitigation assumptions.

Finally, the possibility of quantum advantage could be restricted not only by the finite shot budget but also by the allowed energy consumption.
 Although thermodynamic constraints have been discussed in fault-tolerant scenarios~\cite{Landi2020,wilde2022,thermoqec2024}, the thermodynamic cost of QEM remains largely unexplored~\cite{Bedingham_2016,campbell2025}.
Since some QEM methods, such as zero-noise extrapolation~\cite{temme_error_2017} and virtual distillation~\cite{koczor_exponential_2020, huggins_virtual_2020}, would also incur a quantum overhead, another practical question is therefore which physical cost is required for quantum advantage, in addition to whether QEM increases the success probability.

\section{Acknowledgments}

The authors acknowledge useful feedback from Elham Kashefi and Ra\'ul Garc\'ia-Patr\'on, as well as discussion with Kristan Temme.
B.Y. received funding from the ANR research grants ANR-21-CE47-0014 (SecNISQ), ANR-22-PNCQ-0002 (HQI). 
K.H, M.D. and P.B. were supported by the EPSRC-funded project Benchmarking Quantum Advantage, with reference EP/Y004418/1.
P.B. was also supported by the Quantum Advantage Pathfinder (QAP) programme, with grant reference EP/X026167/1. K.H was also supported by the EPSRC Impact Accelerator Account (IAA).

\section{Data availability statement}
The code used to generate the presented results is publicly available on GitHub \cite{github_link_QEM_aware_QAdv}. 

\appendix

\section{Probabilistic Error Cancellation (PEC)\label{app:pec}}

\begin{figure}[htbp]
    \centering
    \includegraphics[width=\linewidth]{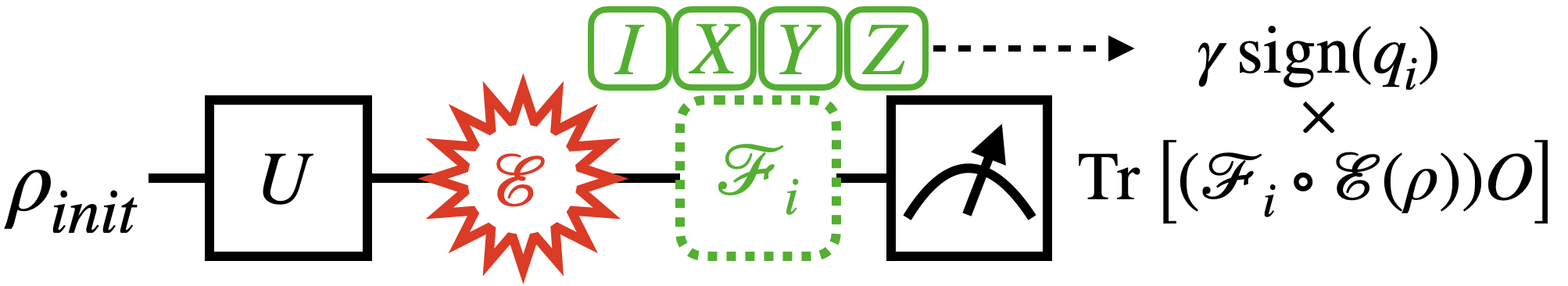}
    \caption{
        \textbf{Schematic illustration of the PEC process.}
    }
    \label{fig:pec}
\end{figure}

\par
The probabilistic error cancellation (PEC) method~\cite{temme_error_2017, endo_practical_2018} restores the noise-free expectation value of an observable by, after having characterized the noise, ``virtually'' applying its inverse 
through a quantum-classical hybrid procedure. We focus here on the latter step.

\par
Given a CPTP channel $\mathcal{E}$ as \textcolor{red}{a} noise model, its inverse $\mathcal{E}^{-1}$ may not be directly applicable through solely a quantum process since it may not be CPTP.
The PEC method overcomes this restriction by decomposing $\mathcal{E}^{-1}$ into $\displaystyle \mathcal{E}^{-1} = \sum_{i=1}^{\mathcal{K}}q_{i}\mathcal{F}_{i}$, where $\{\mathcal{F}_{i}\}_{i=1}^{\mathcal{K}}$ is a set of CPTP maps and $\{q_{i}\}_{i=1}^{\mathcal{K}}$ is a quasi-probability distribution (QPD).
This converts the expectation value of an observable $O$ for a state $\rho$ to the following form,
\begin{equation}
\begin{split}
    \operatorname{Tr}\left[\rho O\right]
    &= \operatorname{Tr}\left[\left(\mathcal{E}^{-1} \circ \mathcal{E}\left(\rho\right)\right) O\right] \\
    &= \sum_{i=1}^{\mathcal{K}} q_{i} \operatorname{Tr}\left[\left(\mathcal{F}_{i} \circ \mathcal{E}\left(\rho\right)\right) O\right] \\
    &= \gamma \sum_{i=1}^{\mathcal{K}} \operatorname{sign}\left(q_{i}\right) p_{i} \operatorname{Tr}\left[\left(\mathcal{F}_{i} \circ \mathcal{E}\left(\rho\right)\right) O\right],
\end{split}
\end{equation}
where $\displaystyle \gamma = \sum_{i=1}^{\mathcal{K}} \left|q_{i}\right|$ is the scaling factor so that $\{p_{i}\}_{i=1}^{\mathcal{K}}$ with elements $\displaystyle p_{i} = \frac{|q_{i}|}{\gamma}$ becomes a valid probability distribution. Since the QPD that can be used is not unique, the associated scaling factor $\gamma$ varies.
In every circuit execution the quantum device computes $\operatorname{Tr}\left[\left(\mathcal{F}_{i} \circ \mathcal{E}\left(\rho\right)\right) O\right]$ with $\mathcal{F}_{i}$ chosen from the Monte-Carlo sampling according to the probability distribution $\{p_{i}\}_{i=1}^{\mathcal{K}}$.
The error-mitigated estimate of $\operatorname{Tr}\left[\rho O\right]$ is then obtained by averaging the whole measurement results with the weight $\gamma \operatorname{sign}\left(q_{i}\right)$ classically.
The schematic illustration of PEC is depicted in Fig.~\ref{fig:pec}.
It is well known that the additional sampling cost of PEC to keep the estimation variance scales with $O(\gamma^{2})$, which is typically a weak-exponential function of the number and size of virtual noise inversions.

\section{Properties of normal distributions \label{app:gaussian_distribs}}

Let $X \sim \mathcal{N}(\mu, \sigma^2)$ be a random variable drawn according to a normal distribution of mean $\mu$ and standard deviation $\sigma$ (namely, the probability density function -- PDF -- associated to $X$ reads $f(x)= \frac{1}{\sqrt{2\pi \sigma^2}}e^{-\frac{(x-\mu)^2}{2\sigma^2}}$). Then the cumulative distribution function (CDF) reads

\begin{align}
    g(x) &= \int_{-\infty}^{x} dt f(t) \\
    &= \int_{-\infty}^{x} dt  \frac{1}{\sqrt{2\pi \sigma^2}}e^{-\frac{(t-\mu)^2}{2\sigma^2}} \\
    &= \frac{1}{2}\left( 1 + \mathrm{erf}\left(\frac{x-\mu}{\sigma \sqrt{2}}\right) \right)
\end{align}

where $\mathrm{erf}$ refers to the error function 

\begin{equation}
    \mathrm{erf}(x) = \frac{1}{\sqrt{\pi}}\int_{-x}^{+x} e^{-t^2}dt.
\end{equation}

This function admits no analytical form, but numerical evaluations are known. As a consequence, the weight of the tail of the normal distribution that is defined by the cut-off value $x_{\mathrm{max}}$ is

\begin{equation}
\label{eq:tail_size_normal_left}
   \mathbb{P}(X \geq x_{\mathrm{max}}) = \frac{1}{2}\left( 1 - \mathrm{erf}\left(\frac{x_{\mathrm{max}}-\mu}{\sigma \sqrt{2}}\right) \right).
\end{equation}
whereas similarly
\begin{equation}
\label{eq:tail_size_normal_right}
   \mathbb{P}(X \leq x_{\mathrm{min}}) = \frac{1}{2}\left( 1 + \mathrm{erf}\left(\frac{x_{\mathrm{min}}-\mu}{\sigma \sqrt{2}}\right) \right).
\end{equation}

Normal distributions frequently arise due to the central limit theorem: the sample average converges in probability law towards a normal distribution, namely if we have $\bar{X}_N = \frac{\sum_i X_i}{N}$ where each sample $X_i$ is drawn from the same distribution with mean $\mu$ and variance $\sigma^2 < \infty$, for large $N$ we have

\begin{equation}
    \sqrt{N} (\bar{X}_N - \mu) \xrightarrow{\text{distrib.}} \mathcal{N}(0, \sigma^2).
\end{equation}
In other words, $\bar{X}_N$ admits as a PDF the function $h(x)=\frac{1}{\sqrt{2\pi \sigma^2 N}}e^{-\frac{(x-\mu)^2}{2\sigma^2}}$.

\section{Effect of erroneous centering of the normal distribution\label{app:effect_error_centering}}

In the proposed method, we make the assumption that the candidate circuit is capable, in the absence of noise, of exhibiting the target ground state energy $E_0$. Since PEC is assumed in the main text to lead to a bias-free estimator, this means we consider a distribution whose mean is $E_0$. Since we do not know $E_0$, we assume it lies in the middle $(E_-+E_+)/2$ of the interval $[E_-, E_+]$ in order to obtain an operational way to assess the probability of succeeding in beating classical methods, given by Eq.~\ref{eq:P_tilde} in the main text. The proposed metric relies further on the assumption of a normal distribution, and is based on measuring the well-defined weight of a normal distribution's tails. In this Appendix, we are concerned about whether setting $E_0 \leftarrow (E_-+E_+)/2$ is a reasonable working assumption.

\begin{figure}
    \centering
    \includegraphics[width=0.7\linewidth]{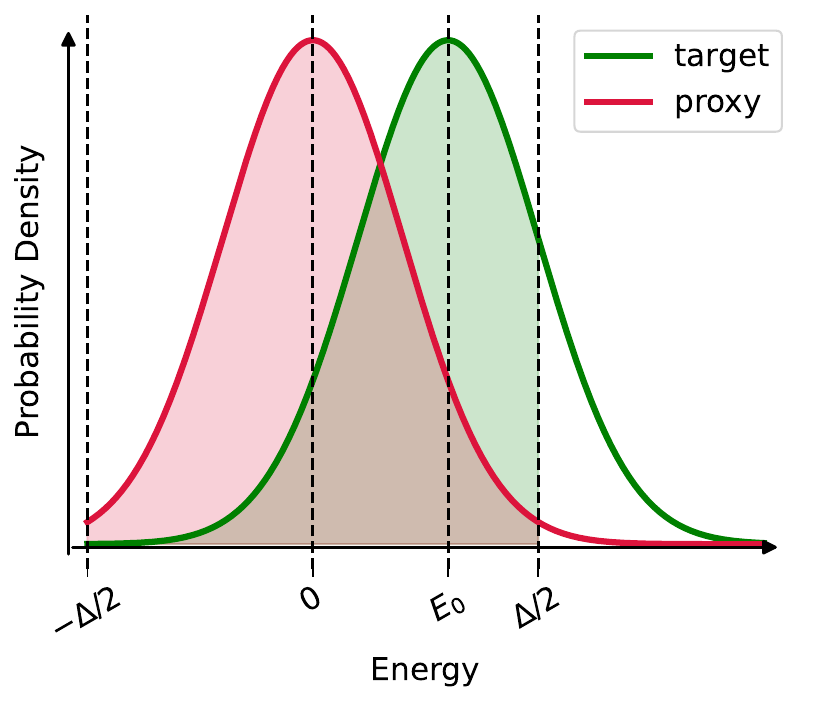}
    \caption{\textbf{Illustration of the study of the accuracy of the proxy $\widetilde{\mathbb{P}}_{\mathrm{success}}$ to the success probability.} We compare $\widetilde{\mathbb{P}}_{\mathrm{success}}$ (red-shaded area) obtained by assuming a Gaussian distribution of standard deviation $\sigma$ and mean 0 with the (actual) success probability $\mathbb{P}_{\mathrm{success}}$ (green-shaded area) obtained for a Gaussian distribution of standard deviation $\sigma$ and mean $E_0$. We vary the relative shift $E_0/(\Delta/2)$ and the relative width of the distributions, $\sigma/\Delta$.}
    \label{fig:study_err_center}
\end{figure}

To this end, we devised the following toy study. We consider an interval $[-\Delta/2, +\Delta/2]$ and a normal distribution of standard deviation $\sigma$ and mean $E_0 \in [-\Delta/2, +\Delta/2]$. Since the method is invariant under a global shift, we can consider this distribution to be a synthetic version of the distribution of energy as studied in the benchmark, and estimate the associated success probability $\mathbb{P}_{\mathrm{success}}$ upon measuring the weight of the tails falling outside the interval $[-\Delta/2, +\Delta/2]$. We compare the obtained value with that of the proxy $\widetilde{\mathbb{P}}_{\mathrm{success}}$ leveraged in the benchmarking strategy set forth, namely here the success probability associated with the gaussian distribution of standard deviation $\sigma$ but mean 0, which is a shifted version of the previous distribution. Without loss of generality, we can assume here $E_0>0$ and focus on the $[0, \Delta/2]$ part of the interval as the 0-mean distribution is symmetric around 0. The approach is illustrated on Fig.~\ref{fig:study_err_center}.

We vary both the relative location of the actual target energy, $E_0/(\Delta/2)$, and the relative width of the distribution $\sigma/\Delta$, with the following principle in mind: the more off-center $E_0$ ($E_0/(\Delta/2) \rightarrow 1$) is, the worse the approximation provided by the proxy value becomes; meanwhile, this is to be put in perspective with the relative width of the distribution, as a highly-peaked distribution ($\sigma/\Delta \rightarrow 0$) will have a probability weight outside the interval that depends less on where it is centered.

\begin{figure}
    \centering
    \includegraphics[width=0.9\linewidth]{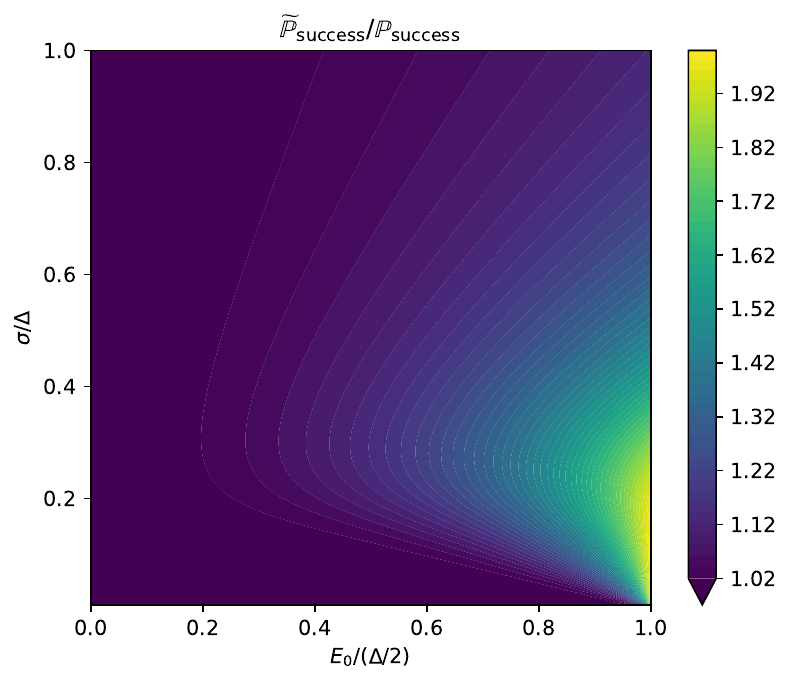}
    \caption{\textbf{Relative error on $\widetilde{\mathbb{P}}_{\mathrm{success}}$ stemming from an erroneous centering of the distribution.}}
    \label{fig:effect_bad_centering}
\end{figure}

We validated this intuition with the diagram provided in Fig.~\ref{fig:effect_bad_centering}. We observe that for sufficiently peaked distributions ($\sigma/\Delta < 0.1$), on a wide range of values for $E_0/(\Delta/2)$ (and as such, a wide range of magnitudes in term of erroneous centering of the distribution), the estimated success probability is overestimated by no more than 12\%. Only in the extreme case where $E_0$ actually lies very close to one side of the interval is the success probability overestimated as being about twice as high at it actually is. In the unlikely scenario where either the known upper or the lower bound to the target energy is a very good approximation whereas the other one is poor, the overestimation is impactful and the method is questionable. However, in a more realistic scenario, one can for instance safely settle for mitigating the effect of erroneously centering the considered energy distribution on the estimation of the success probability by considering a more conservative estimate to the success probability $0.9 \times \widetilde{\mathbb{P}}_{\mathrm{success}}$.

\section{Bias induced by layerwise global depolarizing noise\label{app:bias_depol}}

\par
Consider a global depolarizing noise channel with depolarizing probability $P_D$, of the form 
\begin{equation}
\begin{split}
    \mathcal{E}_D(\rho) = (1-P_D)\rho + P_D\frac{I}{d}.
\end{split}
\end{equation}
We consider $\rho_0$ the ground state of $H$, verifying $\operatorname{Tr}\left[\rho_{0} H\right] = E_0.$

Then, the biased energy of the noisy state corresponding to the ground state affected by $\mathcal{E}_D$ is
\begin{equation}
\begin{split}
    E_\text{noisy}
    &= \operatorname{Tr}\left[\mathcal{E}_D\left(\rho_{0}\right) H\right] \\
    &= (1-P_D)\operatorname{Tr}\left[\rho_{0} H\right] + \frac{P_D}{d}\operatorname{Tr}\left[H\right] \\
    &= (1-P_D)E_{0} + \frac{P_D}{d}\operatorname{Tr}\left[H\right]. \\
\end{split}
\end{equation}

For a $D$-layer quantum circuit under global depolarizing noise with layerwise global depolarizing probability $P$, each global depolarizing noise channel can be pushed to the end of the circuit and combined into a global depolarizing channel with probability $P_D = 1 - (1-P)^D$. We get:
\begin{equation}
    E_\text{noisy} = (1-P)^D E_{0} + \frac{1 - (1-P)^D}{d}\operatorname{Tr}\left[H\right]. \label{eq:E_noisy}
\end{equation}

Thus, we observe that the scale of the energy $E_\text{noisy}$ biased by $D$ layers of global depolarizing noise with probability $P$ is the linear combination of the true energy $E_{0}$ rescaled by a factor $(1-P)^D$ times and the residual term $\displaystyle \frac{1 - (1-P)^D}{d} \operatorname{Tr}\left[H\right]$.

\end{document}